\documentstyle[11pt,epsfig]{article}

\begin{document}

\renewcommand{\thefootnote}{\fnsymbol{footnote}}

\sloppy

\newcommand{\rp}{\right)}

\newcommand{\lp}{\left(}

\newcommand \be  {\begin{equation}}

\newcommand \bea {\begin{eqnarray}}

\newcommand \ee  {\end{equation}}

\newcommand \eea {\end{eqnarray}}

\title{Predicting Financial Crashes Using Discrete Scale Invariance}

\thispagestyle{empty}

\author{Anders Johansen$^1$, Didier Sornette$^{1,2,3}$ and Olivier Ledoit$^4$\\
$^1$ Institute of Geophysics and
Planetary Physics\\ University of California, Los Angeles, California 90095\\
$^2$ Department of Earth and Space Science\\
University of California, Los Angeles, California 90095\\
$^3$ Laboratoire de Physique de la Mati\`{e}re Condens\'{e}e\\ CNRS UMR6622 and
Universit\'{e} de Nice-Sophia Antipolis\\ B.P. 71, Parc
Valrose, 06108 Nice Cedex 2, France \\
$^4$ Anderson Graduate School of Management\\
University of California, Box
90095-1481, 110 Westwood Plaza\\Los Angeles CA 90095-1481\\}

\thispagestyle{empty}

\maketitle

\vskip 2cm

\begin{abstract}
We present a synthesis of all the available empirical evidence in the light
of recent theoretical developments for the existence of characteristic
log-periodic signatures of growing bubbles in a variety of markets
including 8 unrelated crashes from 1929 to 1998 on stock markets as
diverse as the US, Hong-Kong or the Russian market and on currencies.
To our knowledge, no major financial crash preceded by an extended bubble
has occurred in the past 2 decades without exhibiting such log-periodic
signatures.
\end{abstract}

\newpage
\pagenumbering{arabic}

\section{Introduction}

Two recent papers \cite{JS98.2,JLS} have presented increasing evidence
on the Oct.~1929, Oct.~1987, Hong-Kong Oct.~1987 crashes, on the
Aug.~1998 global market events and on the 1985 Forex event on the US dollar,
for the following hypothesis \cite{SJB96} describing stock market crashes\,:
we propose that they are caused by a slow build-up of long-range time
correlations reflecting those between traders leading to a collapse of the
stock market in one critical instant. This build-up manifest itself as an
over-all power law acceleration in the price decorated by  ``log-periodic''
precursors. Here, ``log-periodicity'' refers to a sequence of
oscillations with progressively shorter cycles of a period decaying according
to a geometrical series. In addition, extensive statistical tests have been
performed \cite{JLS,Thesis} to show that the reported ``log-periodic''
structures essentially never occurred in $\approx 10^5$ years
of synthetic trading following a ``classical'' time-series model,
the GARCH(1,1) model with student-t statistics, often used as a benchmark in
academic circles as well as by practitioners.
Thus, the null hypothesis that log-periodicity could result simply from random
fluctuations is strongly rejected.

From a theoretical view point, we have proposed
a rational expectation model of bubbles and crashes, which has
two main components\,: (1) We assume that
a crash may be caused by {\em local} self-reinforcing imitation processes
between noise traders that can be quantified by the theory of critical
phenomena developed in the Physical Sciences\,; (2) We allow
for a remuneration of the risk of a crash by
a higher rate of growth of the bubble, which reflects that
the crash is not a certain deterministic outcome of the bubble and, as
a consequence, it remains rational for traders to remain invested provided
they are suitably compensated.

Since our initial proposition \cite{SJB96}, several works have extended the
empirical investigation
\cite{FF96,97crash,Gluzman,Van1,Van2,manisfesto,Drozdz,antibulle} as well as
the theoretical framework of critical  phenomena \cite{SJ97,SJ98}.
However, a recent article of opinion \cite{Laloux} and a short assessment note
\cite{Ilinski} have raised concerns about the reality of the log-periodic
signatures in bubbles preceding crashes and the relevance of the concept of
critical phenomena to financial crashes in general.

The purpose of the present paper is primarily to present a complete up-to-date
synthesis of the available evidence containing all previously reported cases
together with four new cases which include the correction of the US dollar
against the Canadian dollar and the Japanese
Yen starting in Aug. 1998, as well as the bubble on the Russian market
and its ensuing collapse in 1997-98. We are able to show
a remarkable {\it universality} of the
results for all events, with approximately the same value of
the fundamental scaling ratio $\lambda$
characterising the log-periodic signatures. We also address briefly
the concern raised by Ilinski \cite{Ilinski} about the validity of our
model \cite{JS98.2,JLS}. Considering the robustness of the results
presented here,
our analysis opens the road to financial crash forecasting (see for example
footnote [12] of \cite{StauSor} and \cite{Van2}).

A similar analysis of log-periodicity for ``anti-bubbles'' have been used to
issue such a forecast in Dec. 1998 for the recovery of the Nikkei in 1999
\cite{antibulle}. At the time of writing (May 1999), the forecast, performed
at a time when the Nikkei was at its lowest, seems to have correctly captured
the change of regime and the overall upward trend since the beginning of
this year.

\section{Theoretical framework}

In order to put the empirical analysis in perspective, it is useful to
recapitulate
the main features of the model \cite{JS98.2,JLS}. This will also allows us
to stress its robustness
with respect to other choices in the model construction and address the
criticism made by Ilinski \cite{Ilinski}.

\subsection{Ingredients}

\begin{itemize}

\item Our key assumption is
that a crash may be caused by {\em local} self-reinforcing imitation
between traders \cite{Shiller1,Shiller2,Censi,Cutler,Keim,Nelson}.
This self-reinforcing imitation process leads to the blossoming of a bubble. If
the tendency for traders to ``imitate'' their ``friends'' increases
up to a certain point called the ``critical'' point, many traders may
place the same order (sell) at the same time, thus causing a crash. The
interplay between the progressive strengthening of imitation and the ubiquity
of noise requires a probabilistic description\,: A crash is {\it not} a certain
outcome of the bubble but
can be characterised by its hazard rate $h(t)$, {\it i.e.}, the probability per
unit time that the crash will happen in the next instant provided it has not
happened yet.

\item Since the crash is not a certain deterministic outcome of the bubble, it
remains rational for traders to remain invested provided they are
compensated by a higher rate of growth of the bubble for taking the risk of
a crash, because there is a finite probability of  ``landing smoothly'',
{\it i.e.}, of attaining the end of the bubble without crash. In this model,
the ability to predict the critical date is perfectly consistent with the
behaviour of the rational agents: They all know this date, the crash may
happen anyway, and they are unable to make any abnormal risk-adjusted
profits by using this information.

\end{itemize}

Our model distinguishes between the end of the bubble and the time of the
crash\,: The rational expectation constraint has the specific implication that
the date of the crash must have some degree of randomness. The theoretical
death of the bubble is not the time of the crash because the crash could happen
at any time before, even though this is not very likely. The death
of the bubble is only the most probable time of the crash.

The model does not impose any constraint on the amplitude of the crash.
If we assume it proportional to the current price level, then the natural
variable is the logarithm of the price. If instead, we assume that the crash
amplitude is a finite fraction of the gain observed during the bubble, then
the natural variable is the price itself \cite{JS98.2}.
We are aware that the standard economical proxy is the logarithm of the
price and not the price itself, since only relative variations should play
a role. However, as we shall see below, different price dynamics gives both
possibilities.

In the stylised framework of a purely speculative asset that pays no
dividends and in which we
ignore the interest rate, risk aversion, information asymmetry,
and the market-clearing condition,
rational expectations are simply equivalent to the familiar martingale
hypothesis:

\be \label{eq:martingale}
{\rm for~all}~~ t'>t ~~~~{\rm E}_t[p(t')] = p(t)~,
\ee
where $p(t)$ denotes the price of the asset at time $t$ and E$_t[\cdot]$
denotes the expectation conditional on information revealed up to time $t$.
In the sequel, we will relax the hypothesis of no risk aversion and show
that the predictions of the model are very robust. This is important as
risk aversion is often considered as a key ingredient controlling the
crash dynamics and the time-dependence of the
risk perception by market participants also controls in part the pre-crash
dynamics.

We model the occurrence of a crash as a jump process $j$ whose value is zero
before the crash and one afterwards.
The random nature of the crash occurrence
is modeled by the cumulative distribution function $Q(t)$ of the time of the
crash, the probability density function
$q(t) = dQ/dt$ and the hazard rate $h(t) = q(t)/[1-Q(t)]$. The hazard
rate is the probability
per unit of time that the crash will happen in the next instant provided
it has not happened yet.

Assume for simplicity that, during a crash, the price drops
by a fixed percentage $\kappa\in(0,1)$, say between $20$ and $30\%$ of the
price increase above a reference value $p_1$.
Then, the dynamics of the asset price before the crash are given by:

\be \label{eq:crash}
dp = \mu(t)\,p(t)\,dt-\kappa [p(t)-p_1] dj~,
\ee
where $j$ denotes a jump process whose value is zero before the crash and one
afterwards.
Taking ${\rm E}_t[dp]$ and using the fact that ${\rm E}_t[dj] = h(t) dt$
leads to
\be \label{hfjqklq}
\mu(t) p(t) = \kappa [p(t)-p_1] h(t)~.
\ee
In words, if the crash hazard rate $h(t)$ increases, the return $\mu$
increases to compensate the traders
for the increasing risk. Plugging (\ref{hfjqklq}) into
(\ref{eq:crash}), we obtain a ordinary differential equation. For
$p(t) - p(t_0) < p(t_0) - p_1$, its solution is
\be \label{eq:price}
p(t) \approx p(t_0) +  \kappa [p(t_0) - p_1]~\int_{t_0}^t h(t') dt'
\qquad\mbox{before the crash}.
\ee
The integral $\int_{t_0}^t h(t') dt'$ is the cumulative probability of a crash
until time $t$. We have found that this regime, where the price itself is the
relevant observable, applies to the relatively short time scales of
approximately two to three years prior to the crash studied here.

If instead the price drops by a fixed percentage $\kappa\in(0,1)$ of the
price, the dynamics of the asset price before the crash is given by
\be \label{eq:crash1}
dp = \mu(t)\,p(t)\,dt-\kappa p(t)dj~.
\ee
We then get
\be
{\rm E}_t[dp] = \mu(t)p(t)dt-\kappa p(t)h(t)dt = 0~,
\ee
which yields\,:
\be
\mu(t) = \kappa h(t)~.
\label{hkllmqmlqm}
\ee
and the corresponding equation for the price is\,:
\be
\label{eq:price1}
\log\left[\frac{p(t)}{p(t_0)}\right] = \kappa\int_{t_0}^t  h(t')dt'
\qquad\mbox{before the crash}.
\ee
This gives the logarithm of the price as the relevant observable. It has
successfully been applied to the 1929 and 1987 Wall Street
crashes up to about $7.5$ years prior to the crash \cite{SJ97,JLS}.

The higher the probability of a crash, the faster the price must increase
(conditional on having no crash) in order to satisfy the martingale condition.
Intuitively, investors must be compensated by a higher return
in order to be induced to hold an asset that might crash. This is the only
effect that we wish to capture in this part of the model. This effect is
fairly
standard and it was pointed out earlier in a closely related model of bubbles
and crashes under rational expectations by Blanchard (\cite{Blanchard}, top
of p.389). It
may go against the naive preconception that price is adversely affected by
the probability of the crash, but our result is the only one consistent with
rational expectations. Notice that price is driven by the hazard rate of crash
$h(t)$.

Ilinski \cite{Ilinski} raises the concern that the martingale condition
(\ref{eq:martingale})
leads to a model which ``assumes a zero return as the best prediction for
the market.'' He
continues\,: ``No need to say that this is not what one expects from
a perfect model of market bubble! Buying shares, traders expect the price
to rise and
it is reflected (or caused) by their prediction model. They support the bubble
and the bubble support them!''.

In other words, Ilinski (1999) criticises a key economic element of our model
\,: Market rationality. We have captured this by assuming that the market level
is expected to stay constant as written in equation (\ref{eq:martingale}).
Ilinski (1999) claims that this equation (\ref{eq:martingale})
is wrong because the market level does not stay constant in
a bubble\,: It rises, almost by definition.

This misunderstanding addresses a rather subtle point of the model and
stems from the difference between two different types of returns\,:

\begin{enumerate}
\item The unconditional return is indeed zero as seen from
(\ref{eq:martingale}) and reflects the fair game condition.

\item The conditional return, conditioned upon no crash occurring between
time $t$ and time $t'$, is non-zero and is given by equations (\ref{hfjqklq})
and (\ref{hkllmqmlqm}), respectively. If the crash hazard rate is increasing
with time, the conditional return will be accelerating precisely because the
crash becomes more probable and the investors need to be remunerated for their
higher risk.
\end{enumerate}

Thus, the expectation which remains constant in equation (\ref{eq:martingale})
takes into account the probability
that the market {\em may} crash. Therefore, {\em conditionally}
on staying in the bubble (no crash yet), the market must rationally
rise to compensate buyers for having taken the risk that the market {\em
could} have crashed.

The market price reflects the equilibrium between the greed of buyers
who hope the bubble will inflate and the fear of sellers that it may crash.
A bubble that goes up is just one that could have crashed but did not.
Our model is well summarised by borrowing the words of another economist\,:
``(...) the higher probability of a crash leads to an acceleration
of [the market price] while the bubble lasts.''
Interestingly, this citation is culled from the very same article by
Blanchard \cite{Blanchard} that Ilinski \cite{Ilinski} cites as an alternative
model more realistic than ours. We see that this is in fact more
of an endorsement than an alternative.

To go into details, Blanchard's model \cite{Blanchard} is slightly more
general because it incorporates risk aversion through\,:
\begin{equation}
\nu E_t[p(t')]= p(t)~,
\end{equation}
where $\nu\in(0,1]$ is an appropriate discount factor. We have been aware
of this since the beginning, and the only reason why we did not take it into
account is that it obviously makes {\em no difference} in our
substantive predictions (log-periodic oscillations and power law acceleration),
as long as $\nu$ remains bounded away from zero and infinity.

Another way to incorporate risk aversion into our model is to say that
the probability of a crash in the next instant is perceived by traders
as being $K$ times bigger than it objectively is. This amounts to
multiplying our hazard rate $h(t)$ by $K$, and once again this makes
no substantive difference as long as $K$ is bounded away from zero and
infinity.

The point here is that $\nu$ and $K$ both represent general aversion
of fixed magnitude against a risk.  Risk aversion is a central feature
of economic theory, and it is generally thought to be stable within a
reasonable range, associated with slow-moving secular trends such as
changes in education, social structures and technology.

Ilinski \cite{Ilinski} rightfully points out that risk perceptions are
constantly changing in the course of real-life bubbles, but wrongfully
claims that our model violates this intuition. In our model, risk
perceptions do oscillate dramatically throughout the bubble, even
though subjective aversion to risk remains stable, simply because it is the
{\em objective degree of risk that the bubble may burst} that goes through
wild swings.

For these reasons, the criticisms put forth by Ilinski, far
from making a dent in our economic model, serve instead to show that
it is robust, flexible and intuitive.

\subsection{Crash hazard rate and critical imitation model}

The crash hazard rate $h(t)$ quantifies the probability that a large group
of agents place sell orders simultaneously and create
enough of an imbalance in the order book for market makers to be unable to
absorb the other side without lowering prices substantially.
Most of the time, market agents disagree with one
another, and submit roughly as many buy orders as sell orders (these are all
the times when a crash {\em does not} happen). The key question is to
determine by what
mechanism did they suddenly manage to organise a coordinated sell-off?

We have proposed the following answer \cite{JLS}: All the traders in the
world are organised into a network (of family, friends, colleagues, 
{\it etc.}) and they influence each
other {\em locally} through this network\,: For instance, an active
trader is constantly on the phone exchanging information and opinions with a
set of selected colleagues. In addition, there are indirect interactions
mediated for instance by different parts of the media. Specifically, if I am
directly connected with $k$ other traders, then there are only two forces that
influence my opinion: (a) The opinions of these $k$ people and of the global
information network and (b) an
idiosyncratic signal that I alone generate. Our working hypothesis here is that
agents tend to {\em imitate} the opinions of their connections.
The force (a) will tend to create order, while
force (b) will tend to create disorder. The main story here
is a fight between order and disorder. As far as asset prices are
concerned, a crash happens when order wins (everybody has the same
opinion: Selling), and normal times are when disorder wins (buyers and
sellers disagree with each other and roughly balance each other out).
This mechanism does not require an overarching
coordination mechanism since macro-level coordination can
arise from micro-level imitation and it relies on a
realistic model of how agents form opinions by constant interactions.

Many models of interaction and imitation between traders have been developed.
To make a long story short, the upshot is that the fight between order and
disorder often leads to a regime where order may win. When this occurs,
the bubble ends. Many models (but not all)
undergo this transition in a ``critical'' manner
\cite{Goldenfeld,Dubrulle}\,: The sensitivity of
the market reaction to news or external influences accelerate on the
approach to this transition in a specific way characterized by a power law
divergence
at the critical time $t_c$ of the form $F(t)=(t_c-t)^{-z}$, where $z$ is
called a critical
exponent. This form amounts to the property that
\be
\frac{d\ln F}{d\ln (t_c-t)} = -z      \label{jajalfdlq}
\ee
is a constant, namely that
the behavior of the observable $F$ becomes self-similar close to $t_c$ with
respect to dilation of the distance $t_c-t$ to the critical point at $t_c$.
The symmetry of self-similarity
in the present context refers to the fact that the relative variations
$d\ln F = dF/F$ of the observable with respect to relative variations
$d\ln (t_c-t) = d(t_c-t)/(t_c-t)$ of the time-to-crash are independent of
time $t$, as expressed by the constancy of the exponent $z$.

Accordingly, the crash hazard rate follows a similar dependence, namely

\be
\label{eq:hazard2}
h(t) = B (t_c-t)^{-b}
\ee
where $B$ is a positive constant and $t_c$ is the critical point or the
theoretical date of the bubble death. The exponent $b$ must lie between
$0$ and $1$ for an important economic reason\,: otherwise, the price would
go to infinity when approaching $t_c$ (if the bubble has not crashed yet).

We stress that $t_c$ is {\it not} the time of the crash because the crash
could happen at any time before $t_c$, even though this is not very likely.
$t_c$ is the most probable time of the crash. There
exists a residual probability
\be
1- \int_{t_0}^{t_c} h(t) dt >0
\ee
of attaining the critical date without crash. This residual probability is
crucial for the coherence of the story, because otherwise the whole model
would unravel because rational agents would anticipate the crash.

Plugging equation (\ref{eq:hazard2}) into Equation (\ref{eq:price}) gives the
following law for price:
\be
\label{eq:solution}
p(t) \approx p_c -\frac{\kappa B}{\beta} (t_c-t)^{\beta}
\qquad\mbox{before the crash}.
\ee
where $\beta = 1-b\in(0,1)$ and $p_c$ is the price at the critical
time $t_c$ (conditioned on no crash having been triggered). We
see that the price before the crash also follows a power law with a
finite upper bound $p_c$. The slope of the price,
which is the expected return per unit of time, becomes unbounded as we
approach the critical date $t_c$. This is to compensate for an unbounded
hazard rate approaching $t_c$.

In the contemporary information environment,
one could argue that the media can play a more efficient role in the
creation of
an opinion than the network of ``friends'' and will thus modify if not destroy
the critical nature of the crash resulting from the increasing
cooperativity between
the network of ``friends''. There are two ways to address this issue.
First, consider the situation of a typical trader. Since
all her competitors have access to the same globally shared information,
she does not feel
that this gives her an edge, but only serves to keep her abreast the market
moves. The informations that she consider really valuable are those that she
can share and exchange
with a selected and narrow circle of trusted colleagues. This argument
brings us back to the
model of an imitation network with ``local'' connections. A second argument
is that
the media do nothing but reflect a kind of consensus resulting precisely from
the collective action of all the actors on the market. It is thus a kind of
revelator of the interactions between the traders. In addition, this
revelation of the general consensus and of the global market sentiment
reinforces the effective interaction between the traders.
In the language of the Statistical Physics
of critical points, the media can thus be compared to an effective mean-field
created endogenously and may thus
effectively strengthen the existence of a critical point as described here.

\subsection{Log-periodicity}

The power law dependence (\ref{eq:hazard2}) of the hazard rate is the hallmark
of self-similarity of the market across scales as the critical
time $t_c$ is approached\,: at the critical
point, an ocean of traders who are mostly bearish may have within it several
islands of traders who are mostly bullish, each of which in turns surrounds
lakes of bearish traders with islets of bullish traders; the progression
continues all the way down to the smallest possible scale: a single trader
\cite{Wilson}. Intuitively speaking, critical self-similarity is why
local imitation cascades through the scales into global coordination.
This critical state occurs when local influences
propagate over long distances and the average state of the system becomes
exquisitely sensitive to a small perturbation, {\it i.e.}, different parts of
the system becomes highly correlated.

As we said, scale invariance of a system near its critical point
must be represented by power law dependences of the observables
\cite{Dubrulle}.
Formally, these power laws are the solution of the renormalisation group
equations \cite{Wilson,Goldenfeld}
which describe the correlations between agents at many scales. It turns out
that the most general mathematical solutions of these renormalisation group
equations
are power laws with {\it complex} exponents which as a consequence exhibits
log-periodic corrections to scaling \cite{Revue}.

A straightforward mechanism for these complex exponents to appear
is to define the dynamics on a hierarchical structure displaying
a discrete scale invariance. Schematically, we can think of
the stock market made of actors which differs in
size by many orders of magnitudes ranging from individuals to gigantic
professional investors, such as pension funds. Furthermore, structures at
even higher levels, such as currency influence spheres (US\$, Euro, YEN ...),
exist and with the current globalisation and de-regulation of the market
one may argue that structures on the largest possible scale, {\it i.e.},
the world economy, are beginning to form. This means that the structure
of the financial markets have features which resembles that of hierarchical
systems with ``traders'' on all levels of the market. Of course, this
does not imply that any strict hierarchical structure of the stock market
exists, but there are numerous examples of qualitatively hierarchical
structures in society. Models \cite{JLS,SJ98} of imitative interactions on
hierarchical structures recover the power law behavior (\ref{eq:solution}) and
predict that the critical exponents $b$ and $\beta$ may be complex numbers!

Recently, it has been realized that discrete scale invariance and its
associated complex
exponents may appear ``spontaneously'' in non-hierarchical systems, {\it i.e.},
without the need for
a pre-existing hierarchy (see \cite{Revue} for a review). There are many
dynamical mechanisms
that can operate to produce these complex exponents, such as non-local
geometry, frozen heterogeneity \cite{SaleurSor}, cascades of instabilities,
intermittent amplifications,  (see \cite{antibulle} for such a mechanism
in the context of ``anti-bubbles'' proposed recently for the Japanese stock
market), and so on.

As a consequence, we are led to generalize (\ref{eq:hazard2}) and give
the first order expansion of the general solution for the hazard rate
\be \label{eq:hazard3}
h(t)\approx B (t_c-t)^{-b}
+C(t_c-t)^{-b}\cos[\omega_1 \log(t_c-t)+\psi].
\ee
The crash hazard rate now displays log-periodic oscillations. This can easily
seen to be the generalization of equation (\ref{jajalfdlq}) by taking the
exponent $z$ to be complex with a non-zero imaginary part, since the real part
of $(t_c-t)^{-z+i\omega}$
is $(t_c-t)^{-z}~\cos \left( \omega \ln (t_c-t) \right)$. The evolution
of the price  before the crash and before the critical date is then given by:
\be \label{lppow}
p(t) \approx A_1 + B_1 (t_c-t)^{\beta}
+C_1(t_c-t)^{\beta}\cos\lp\omega_1\log(t_c-t)+\phi_1\rp
\ee
where $\phi_1$ is another phase constant. The key feature is that
accelerating oscillations
appear in the price of the asset before the critical date. The local
maxima of the function are separated by time intervals that tend to zero at the
critical date, and do so in geometric progression such that the ratio of
consecutive time intervals is a constant

\be
\lambda \equiv e^{2 \pi / \omega_1}~.
\label{hklmqmmq}
\ee

This is very useful from an empirical point
of view because such oscillations are much more strikingly visible in actual
data than a simple power law\,: a fit can ``lock in'' on the oscillations which
contain information about the critical date $t_c$.

We note that these log-periodic structure bear some similarity with
patterns classified empirically by chartists and others using methods of
technical analysis, such as ``Elliott waves'' \cite{Elliott} and
``log-spirals'' \cite{spiral}. For instance, a logarithmic spiral in the plane
obeys the equation $r = r_0 ~e^{a \theta}$ in polar coordinate. The
intersections with the $x$ axis occur for $\theta = n \pi$ (recall that
$x = r \cos \theta$) and the corresponding coordinates are $x_n =
(-1)^n~e^{a\pi n}$.
The  intersections with the $y$ axis occur for $\theta = \pi/2 + n \pi$
(recall that $y = r \sin \theta$) and the corresponding coordinates are
$y_n = (-1)~e^{a\pi/2}~~e^{a\pi n}$.  Both series indeed form discrete
geometrical series
with a scaling ratio $\lambda = e^{a\pi}$ similar to (\ref{hklmqmmq}).
We do not need, however, to stress the difference between a pattern recognition
approach \cite{spiral} and our economically motivated approach incorporating
a rational model of imitative agents as well as well-established tools from
statistical physics.

\section{Empirical results}

\subsection{Are large crashes special?} \label{outlier}

Crashes are extreme events. There are two possibilities to describe them\,:

\begin{enumerate}
\item The distribution of returns is stationary and
the extreme events can be extrapolated as lying in its far tail.
Within this point of view, recent works
in finance and insurance have recently investigated the relevance of the
body of theory known as Extreme Value Theory to extreme events and crashes
\cite{Embrechts1,Embrechts2,Embrechts3}.

\item Crashes cannot be accounted for by an extrapolation of the
distribution of smaller events to the regime of extremes
and belong intrinsically to another regime, another distribution,
and are thus outliers.

\end{enumerate}

To test which one of these two alternatives provides the most accurate
description, a statistical analysis of market fluctuations \cite{JS98.1}
has provided significant indications that large crashes may be outliers.
Indeed, it was established that the distribution of draw downs
of the Dow Jones Average daily closing is well
described by an exponential distribution with a decay constant of
about $2\%$. This exponential distribution holds only for draw downs
smaller than
about $15\%$.
In other words, this means that all draw downs of amplitudes of up
to approximately $15\%$ are well-described by the same exponential distribution
with characteristic scale $2\%$. This
characteristic decay constant means that
the probability to observe a draw down larger than $2\%$ is about $37\%$.
Following hypothesis 1 and extrapolating this description to, {\it e.g.},
the 3 largest crashes on Wall Street in this century (1914, 1929 and 1987)
yields a recurrence time of about $50$ centuries for {\it each single} crash.
In reality, the three crashes occurred in less than one century. This
suggests that hypothesis 2 is preferable.
In the following, we use the generic term ``crash'' to refer to the
significant drop of a market occurring over a few days to a few weeks
which follows an extended period of bullish behaviour and signifies the
end of the bubble.

As an additional null-hypothesis, $10.000$ synthetic data sets, each covering
a time-span close to a century hence adding up to about $10^6$ years, have been
generated using a GARCH(1,1) model estimated from the true index
with a t-student distribution with four degrees of freedom \cite{GARCH}.
This model is often used in academic and practitioner circles
as a reasonable description of the market and
includes both the non-stationarity of volatilities and the fat tail nature
of the price returns.
Our analysis \cite{JLS} showed that only two data sets had $3$ draw downs
above $22$\% and none had $4$.
However, $3$ of these $6$ ``crashes'' were preceded by a draw up of comparable
size of the draw down and hence showed an abnormal behaviour not found for real
crashes. This means that in approximately one million years of
``Garch-trading'', with a reset every century, {\it never} did $3$
crashes occur in a single century.

This suggest that different mechanisms may be responsible for large crashes
and that hypothesis 2 is the correct description of crashes. This also raises
the issue whether pre-cursory patterns exists decorating the speculative bubble
preceding the crash.  The point is simply that while the GARCH(1,1) model does
a reasonable job of reproducing fluctuations in ``normal trading'', it cannot
capture the fluctuations connected with large crashes and hence another type
of model is necessary for this special behaviour. Of course, these simulations
do not prove that our model is the correct one, only that one of the
standard models of
the ``industry'' (which makes a reasonable null hypothesis) is utterly
unable to account for the stylized facts associated to large financial crashes.
A better model is thus called for and this is exactly the goal of
the present paper to provide for such a framework.

\subsection{Log-periodic analysis of large crashes}

The numerical procedure of fitting equation (\ref{lppow}), as well as equations
(\ref{1feq}), (\ref{2feq}) and (\ref{3feq}) is a minimisation of the variance
\be \label{var}
Var = \frac{1}{N} \sum_{i=1}^{N} \lp y_i - f\lp t_i\rp \rp^2
\ee
between the $N$ data points $y_i$ and the fit function $f\lp t\rp$ using the
down-hill simplex  algorithm
\cite{Numrec}. Tests using the maximum likelihood method with the student-t
distribution show that our results are not significantly sensitive to the
estimation procedure \cite{Pansy}.
In order to reduce the complexity of the fit, all linear
variables $A,B,C$ are determined by the nonlinear variables by requiring that
a solution has zero derivative of the variance as a function of the linear
variables. This means that equation (\ref{lppow}), is an effective
4-parameter fit with $\beta ,\phi , t_c , \omega$. Further details on our
numerical procedure are given in \cite{JLS,Thesis}.

A remaining question concerns what data interval prior to the crash to fit.
The procedure used was the following. The last point used for all crashes
was the highest value of the price before the crash, the first point used
was the lowest value of the price when the bubble started. For all the cases
discussed here, except the Russian anti-bubble analysed in section 
\ref{rus}\footnote{Here ``lowest'' and ``highest'' is of course interchanged},
this procedure always gave a
convincing result and it was never necessary to change the interval fitted.

\begin{figure}
\begin{center}
\epsfig{file=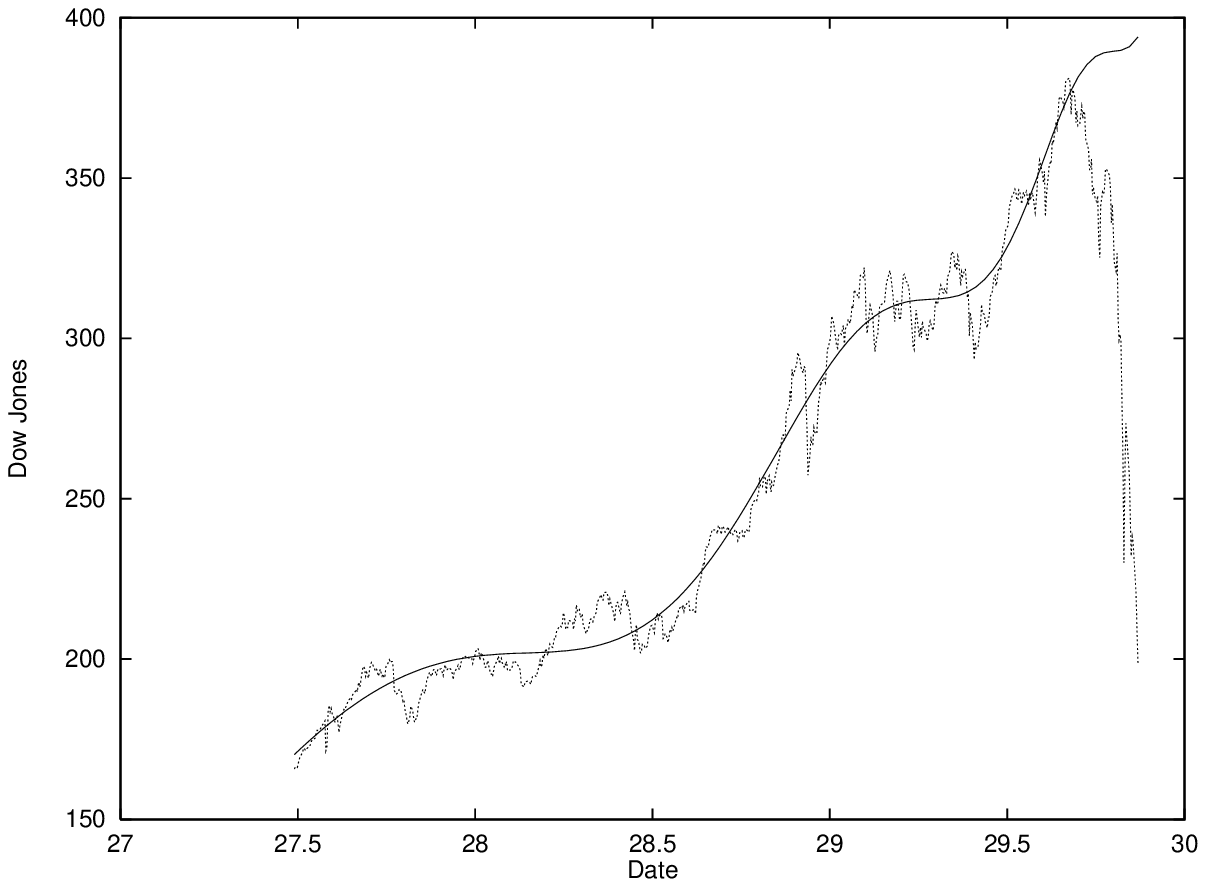,width=10cm}
\caption{\protect\label{ws29}The Dow Jones index prior to the October 1929
crash on Wall Street. The fit is equation (\protect\ref{lppow}) with
$A_1 \approx  571 $, $ B_1 \approx -267 $, $C_1\approx 14.3 $, $\beta\approx
0.45 $, $ t_c \approx  30.22$, $ \phi_1 \approx  1.0 $, $ \omega_1 \approx
7.9$. }
\end{center}
\end{figure}

\begin{figure}
\begin{center}
\epsfig{file=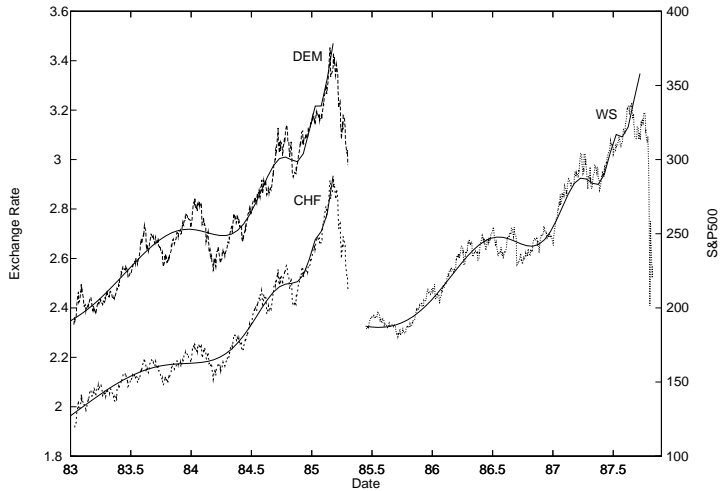,width=10cm}
\caption{\protect\label{forex85ws87} The S\&P 500 US index prior to the
October
1987 crash on Wall Street and the US \$ against DEM and CHF prior to the
collapse mid-85. The fit to the S\&P 500 is equation (\protect\ref{lppow}) with
$A_1\approx  412  $, $ B_1\approx  -165 $, $C_1 \approx   12.2 $, $  \beta
\approx   0.33 $, $t_c \approx   87.74  $, $ \phi_1 \approx   2.0 $, $
\omega_1 \approx
7.4$. The fits to the DM and CHF currencies against the US dollar gives
$A_1\approx  3.88  $, $ B_1\approx  -1.2 $, $ C_1 \approx   0.08 $, $  \beta
\approx   0.28 $, $t_c \approx   85.20  $, $ \phi_1 \approx   -1.2 $, $
\omega_1 \approx   6.0$ and $A_1\approx  3.1  $, $ B_1\approx  -0.86 $, $
C_1 \approx   0.05 $, $  \beta \approx   0.36 $, $
t_c \approx   85.19  $, $ \phi_1 \approx   -0.59 $, $ \omega_1 \approx   5.2$,
respectively.}
\end{center}
\end{figure}

\begin{figure}
\begin{center}
\epsfig{file=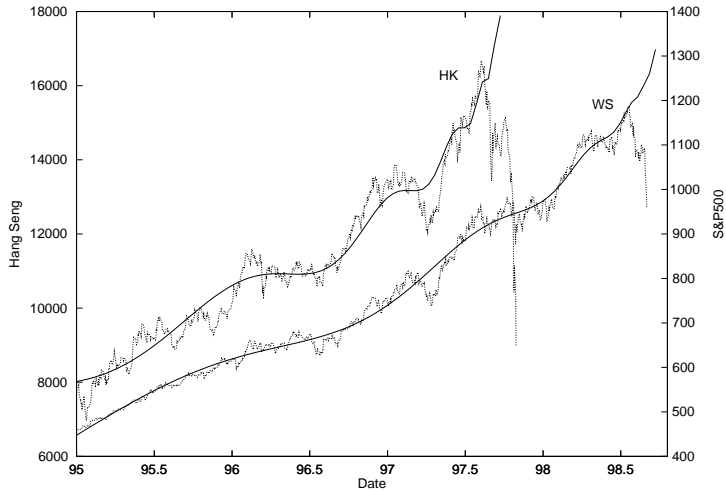,width=10cm}
\caption{\protect\label{hk97ws98}The Hang Seng index prior to the October 1997
crash on the Hong-Kong Stock Exchange and the S\&P 500 stock market index
prior to the
recent crash on Wall Street in August 1998. The fit to the Hang Seng index
is equation (\protect\ref{lppow})
with $A_1 \approx  20077 $, $ B_1\approx  -8241 $, $
\frac{\kappa}{\beta} B_1\approx  -397$, $\beta\approx  0.34 $, $ t_c\approx
97.74 $, $ \phi_1\approx  0.78 $, $ \omega_1\approx  7.5$.
The fit to the S\&P 500 index is equation (\protect\ref{lppow})
with $A_1 \approx  1321$, $ B_1\approx -402 $, $
C_1 \approx  19.7 $, $ \beta \approx  0.60$, $
 t_c\approx  98.72 $, $ \phi_1 \approx  0.75$, $ \omega_1 \approx  6.4$. }
\end{center}
\end{figure}

\begin{figure}
\begin{center}
\epsfig{file=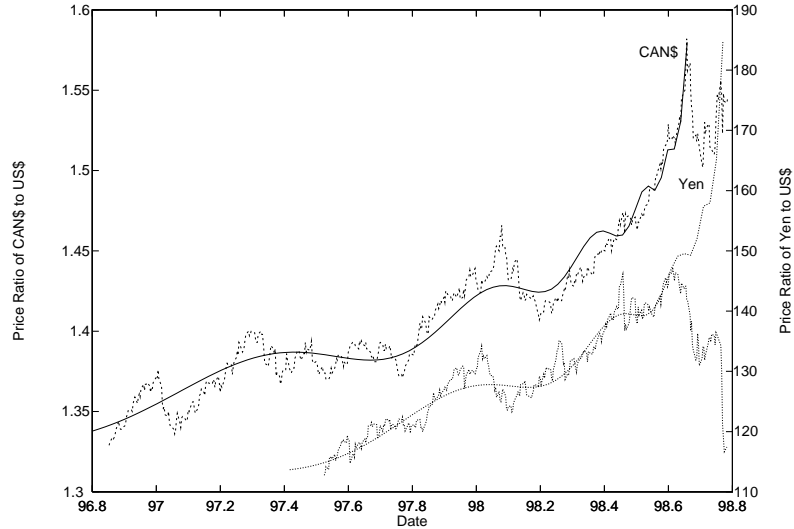,width=10cm}
\caption{\protect\label{forex98}The CAN\$ and YEN currencies against the US
dollar prior to the drop starting in Aug. 1998. The fit with equation
(\protect\ref{lppow}) to the two exchange rates gives $A_1\approx 1.62  $,
$ B_1\approx  -0.22 $, $C_1 \approx   -0.011 $, $  \beta \approx   0.26 $, $
t_c \approx   98.66  $, $ \phi_1 \approx   -0.79 $, $ \omega_1 \approx
8.2$ and $A_1\approx  207  $, $ B_1\approx  -85 $, $C_1 \approx   2.8 $,
$\beta \approx 0.19 $, $ t_c \approx   98.78  $, $ \phi_1 \approx   -1.4 $,
$ \omega_1 \approx  7.2$,
respectively. }
\end{center}
\end{figure}

Figures \ref{ws29} to \ref{forex98} show the behaviour
of the market index prior to the 4 stock market crashes of Oct. 1929 (Wall
Street), Oct. 1987 (Wall Street), Oct. 1997 (Hong-Kong) and  Aug. 1998 (Wall
Street) as well as the collapse of the US\$ against the DEM and CHF in 1985
and against the Canadian dollar and the YEN in 1998. A fit with equation
(\ref{lppow}) is shown as a continuous line for each event. Table \ref{table1}
lists the parameters of the fit to the data. Note the small fluctuations in
the value of the scaling ratio $2.2 \leq \lambda \leq 2.8$ for all data
sets except the CHF. This agreement cannot be accidental and constitutes one
of the key test of our framework.

\begin{table}[h]
\begin{center}
\begin{tabular}{|c|c|c|c|c|c|c|c|c|c|c|c|} \hline
crash & $t_c$ & $t_{max}$ & $t_{min}$ & drop & $\beta$ & $\omega_1$ &
$\lambda$ & $A$ & $B$ & $C$ & $Var$ \\ \hline
1929 (WS)&  $30.22$ & $29.65$ & $29.87$ & $47\%$ & $0.45$ & $7.9$ &
$2.2$ & $571$ & $-267$ & $14.3$ & $56$  \\ \hline
1985 (DEM) &  $85.20$ & $85.15$ & $85.30$ & $14\%$ & $0.28$ & $6.0$ &
$2.8$ & $3.88$ & $1.16$ & $-0.77$ & $0.0028$ \\ \hline
1985 (CHF) &  $85.19$ & $85.18$ & $85.30$ & $15\%$ & $0.36$ & $5.2$ &
$3.4$ & $3.10$ & $-0.86$ & $-0.055$ & $0.0012$ \\ \hline
1987 (WS)&  $87.74$ & $87.65$ & $87.80$ & $30\%$ & $0.33$ & $7.4$ &
$2.3$ & $411$ & $-165$ & $12.2$ & $36$ \\ \hline
1997 (H-K) &  $97.74$ & $97.60$ & $97.82$ & $46\%$ & $0.34$ & $7.5$ &
$2.3$ & $20077$ & $-8241$ & $-397$ & $190360$ \\ \hline
1998 (WS)&  $98.72$ & $98.55$ & $98.67$ & $19\%$ & $0.60$ & $6.4$ &
$2.7$ & $1321$ & $-402$ & $19.7$ & $375$ \\ \hline
1998 (YEN)& $98.78$ & $98.61$ & $98.77$ & $21\%$ & $0.19$ & $7.2$ &
$2.4$ & $207$ & $-84.5$ & $2.78$ & $17$\\ \hline
1998 (CAN\$)&$98.66$& $98.66$ & $98.71$ & $5.1\%$& $0.26$ & $8.2$ &
$2.2$ & $1.62$ & $-0.23$ & $-0.011$ & $0.00024$\\ \hline
\end{tabular}
\end{center}
\caption{\label{tabel}\label{table1} $t_c$ is the critical time
predicted
from the fit of the market index to the equation (\ref{lppow}). The other
parameters of the fit are also shown. The error $Var$ is the variance between
the data and the fit, equation (\protect\ref{var}), and has units $price^2$. 
Each fit is performed up to the time $t_{max}$ at which the market index 
achieved its highest maximum before the crash. $t_{min}$ is the time of the 
lowest point of the market disregarding smaller ``plateaus''. The percentage 
drop is calculated from the total loss from $t_{max}$ to $t_{min}$.}
\end{table}

In order to qualify further the significance of the log-periodic oscillations
in a non-parametric way as well as providing an independent test on the value
of the frequency of the log-periodic oscillations and hence the prefered
scaling ratio $\lambda$, we have eliminated the leading trend
from the price data by the transformation

\be \label{residue}
p\lp t\rp \rightarrow \frac{p\lp t\rp -  A_1 -
B_1(t_c-t)^{\beta}}{C_1(t_c-t)^{\beta}}~ .
\ee

\begin{figure}
\begin{center}
\epsfig{file=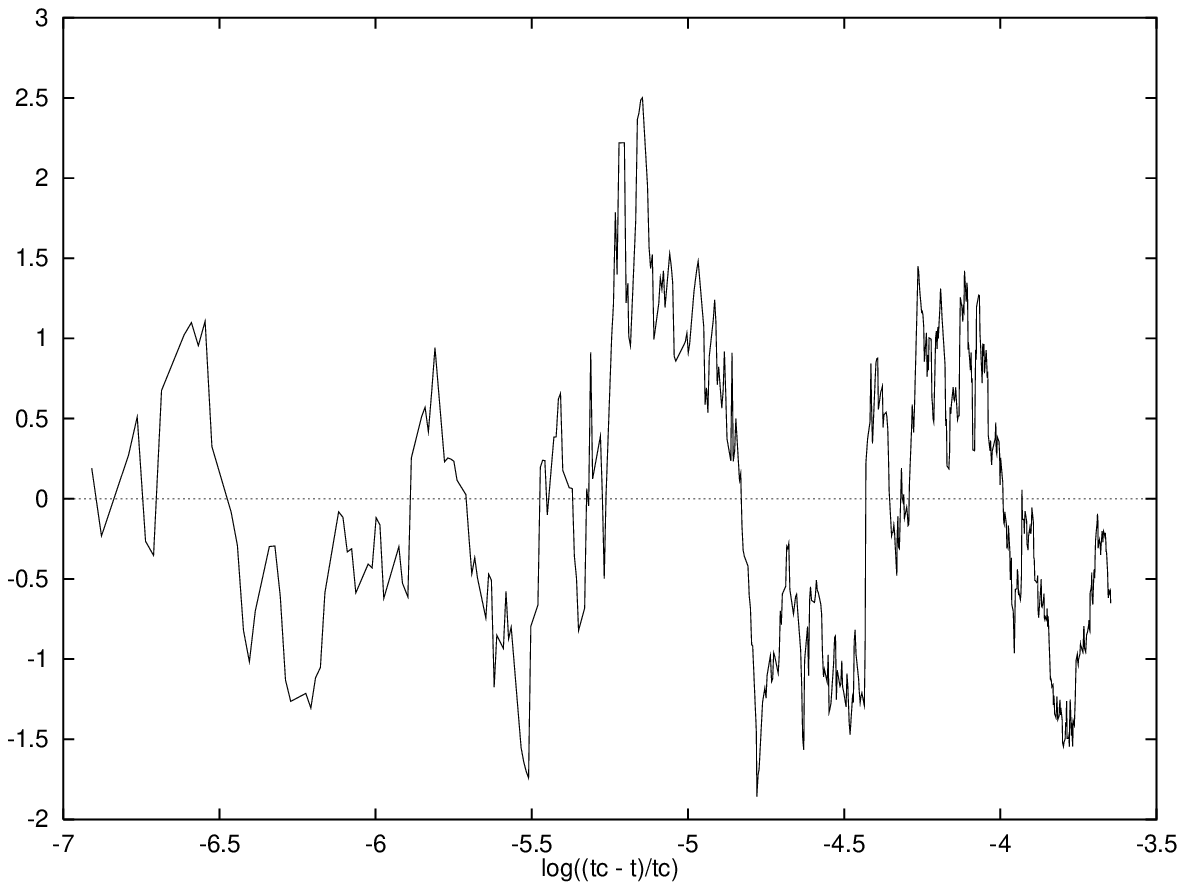,  width=10cm}
\caption{\protect\label{residue87} The residual as defined by the
transformation
(\protect\ref{residue}) as a function of $\log\lp \frac{t_c - t}{t_c}\rp$ for
the 1987 crash.}
\end{center}
\end{figure}

\begin{figure}
\begin{center}
\epsfig{file=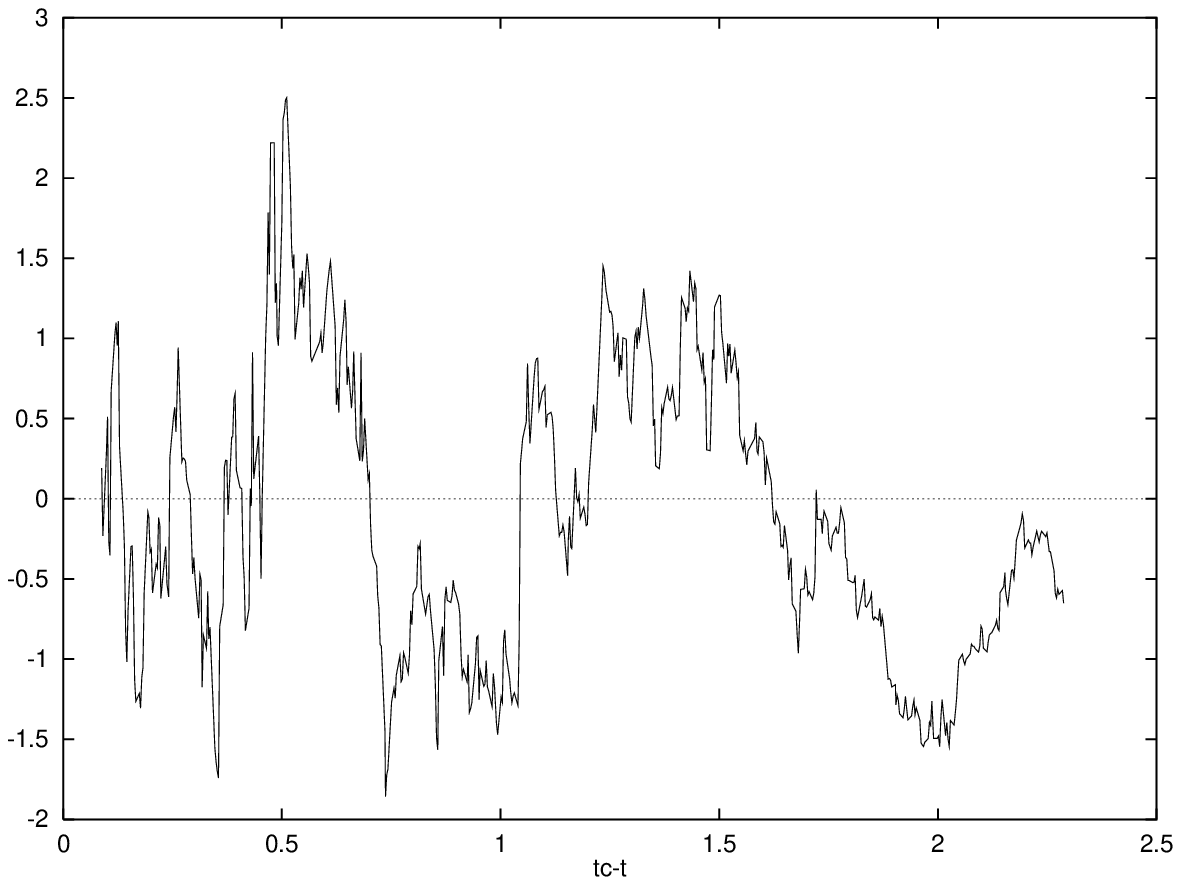,  width=10cm}
\caption{\protect\label{linresidue87} The residual as defined by the
transformation
(\protect\ref{residue}) as a function of $t_c - t$ for the 1987 crash.}
\end{center}
\end{figure}

This transformation should produce a pure $\cos[\omega_1\log(t_c-t)+\phi_1]$
{\it if} equation (\ref{lppow}) was a perfect description. In figure
\ref{residue87}, we show the residual for the bubble prior to the 1987 crash.
We see a very convincing periodic behaviour as a function of
$\log\lp \frac{t_c - t}{t_c}\rp$. In figure \ref{linresidue87}, we see the
corresponding data as a function of the true time to the crash $t_c -t$. 
Looking at
figures \ref{residue87} and \ref{linresidue87}, especially two things strikes
the eye. First. we see that in the latter figure the distance
between consecutive extrema of the data is clearly accelerating, in the
former we clearly see that a transformation to the logarithm of the time to
the crash
exhibits an over-all periodic trend, hence demonstrating a log-periodic
behaviour. Second,
since the detrended data oscillates nicely around zero, the two figures
clearly show that the power law in equation (\ref{lppow}) is an excellent
approximation to the over-all rise in the data. That the residual oscillates
beyond $\pm 1$ only indicates that the cosine used in equation (\ref{lppow})
is not an optimal choice for a periodic function in the present context.
However, it has the advantage of being familiar to the majority as well as
straightforward to implement numerically.

\begin{figure}
\begin{center}
\epsfig{file=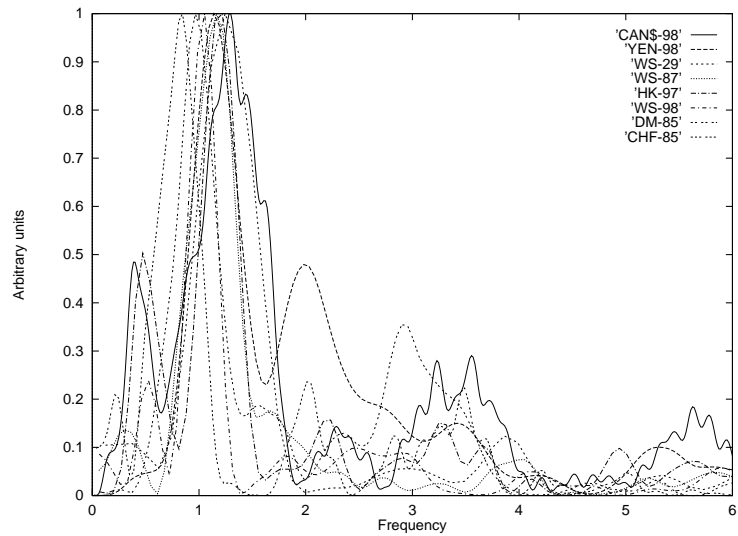,width=10cm}
\caption{\protect\label{allaccfp}The Lomb periodogram for the 1929, 1987
and 1998 crashes and Wall Street, the 1997 crash on the Hong Kong
Stock Exchange, the 1985 US \$ currency crash in 1985 against the DM and CHF
and in 1998 against the YEN and the $5.1\%$ correction against the CAN\$.
For each periodogram, the significance of the peak should be estimated against
the noise level. }
\end{center}
\end{figure}

The significance of this oscillating trend has been further
estimated by performing a spectral analysis, designed to test for the
statistical significance of the log-periodic oscillations. In the goal,
we use a so-called Lomb periodogram \cite{Numrec} rather than the more
standard power spectrum, since the data is not
equidistantly sampled in time. The Lomb periodogram is a local fit of a
cosine (with a phase) using some user chosen range of frequencies. In figure
\ref{allaccfp}, we see a peak around $f \approx 1.1$ for all 8 cases shown
in figures \ref{ws29} to \ref{forex98} corresponding to $\omega_1 = 2\pi f
\approx 7$ in perfect agreement with the results listed in table \ref{table1}.
Since the ``noise'' spectrum is unknown and very likely different for each
crash, we cannot estimate the confidence interval of the peak in the usual
manner \cite{Numrec} and compare the results
for the different crashes. Therefore, only the relative level of the peak
{\it for each separate} periodogram can be taken as a measure of the
significance of the oscillations and the periodograms have hence been
normalised. We note that {\it if} the noise were to have a white Gaussian
distribution, the confidence given by the periodograms increases exponentially
on the value of the abscissa and would be well above $99.99\%$ for all cases
shown \cite{Numrec}. Note also, that the strength of the oscillations is
$\approx 5 - 10\%$ of the leading power law behaviour for all 8 cases
signifying that they cannot be neglected.

To summarize thus far, our spectral analysis demonstrates that
the observed log-periodic oscillations have a very strong power spectrum,
much above noise level. It would be very difficult and much less
parsimonious to account for these structures by another model.

\subsection{Are log-periodic signatures statistically significant?}

In the case of the 1929 and 1987 crashes on Wall Street, log-periodic
oscillations in the index could be identified as long as $\approx 7.5$ years
prior to the crashes \cite{SJ97}.  In order to investigate the significance
of these results, we picked at random fifty  $400$-week intervals in the period
1910 to 1996 of the Dow Jones average and launched the same fitting procedure
as on the time period prior to the 1929 and 1987 crashes on Wall Street. The
results were encouraging. Of 11 fits with a quality (specifically, the
variance of the fit of equation (\ref{2feq}) with $\tau = t_c -t$ to the
data) comparable with
that of the 2 crashes, only six data sets produced values for $\beta$,
$\omega_1$, $\omega_2$ and $T_1$, which were in the same
range. However, all of these fits belonged to the periods prior to the crashes
of 1929, 1962 and 1987. The existence of a ``crash'' in 1962 was before these
results unknown to us and the identification of this crash naturally
strengthens the case. We refer the reader to \cite{JLS} for a presentation of
the best fit obtained for this ``crash''.

We also generated 1000 synthetic data sets of length 400 weeks using the
same GARCH(1,1) model used in section \ref{outlier}. Each of these 1000
data sets was analysed in the same manner as the real crashes. In 66 cases,
the best minima had values similar to the real crashes, but all of these fits,
except two, did not resemble the true stock market index prior to the 1929
and 1987 crashes on Wall Street, the reason primarily being that they only
contained one or two oscillations. However, two fits looked rather much like
that of the 1929 and 1987 crashes \cite{JLS,Thesis}.

This result correspond approximately to the usual 95 \% confidence interval
for {\em one} event.
In contrast, we have here provide 5  examples of log-periodic
signatures before large stock market crashes. We are aware that this is not
proof of log-periodic signatures in bubbly markets, this only a thorough data
analysis can only establish within reasonable doubt, but it provides for a
very good estimation that these signatures are not easily generated
accidentally.

\subsection{Log-periodicity in bearish markets?} \label{decay}

Stock market jargon divide the stock markets trends into either
``bullish'' or ``bearish''. If log-periodic signals, apparently, are
present in ``bullish'' markets, the obvious question to ask is whether
this is also the case in ``bearish'' markets.

\begin{figure}
\begin{center}
\epsfig{file=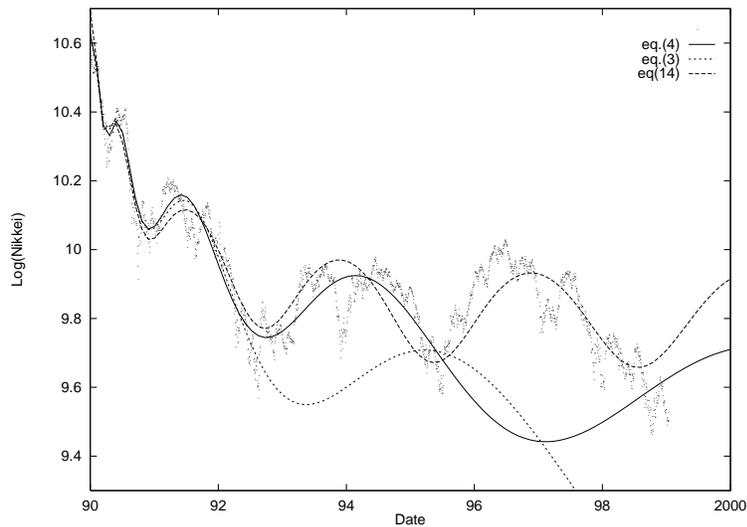,width=10cm}
\caption{\protect\label{decnikkei} Natural logarithm of the Nikkei stock market
index after the start of the decline 1. Jan 1990 until 31 Dec. 1998. The lines
are equation (\protect\ref{1feq}) (dotted line) fitted over an interval of
$\approx 2.6$ years, equation (\protect\ref{2feq}) (continuous line) over
$\approx 5.5$ years and equation (\protect\ref{3feq}) (dashed line) over $9$
years. The parameter values of the first fit of the Nikkei are $A_1 \approx
10.7 , B_1\approx -0.54 , C_1\approx  -0.11 , \beta \approx  0.47 , t_c
\approx  89.99 , \phi_1 \approx -0.86 , \omega_1 \approx  4.9$ for equation
(\protect\ref{1feq}).
The  parameter values of the second fit of the Nikkei are $A_2 \approx
10.8 ,  B_2 \approx  -.70 , C_2 \approx  -0.11 , \beta \approx  0.41 , t_c
\approx  89.97 ,   \phi_2 \approx  0.14 , \omega_1 \approx  4.8 , T_1
\approx 9.5 , \omega_2 \approx 4.9$ for equation (\protect\ref{2feq}).
The third fit uses the entire time interval and is performed by adjusting only
$T_1$, $T_2$,
$\omega_2$ and $\omega_3$, while $\beta$, $t_c$ and $\omega_1$ are
fixed at the values obtained from the previous fit. The values obtained for
these four parameters are $T_1 \approx 4.3$ years, $T_2 \approx 7.8$ years,
$\omega_2 \approx -3.1 $ and $T_2 \approx 23$. Note that
the values obtained for the two time scales $T_1$ and $T_2$
confirms their ranking. This last fit predicts a change of regime and that
the Nikkei should increase in 1999. }
\end{center}
\end{figure}

The most recent example of a genuine long-term depression comes from Japan,
where the Nikkei has decreased by more than $60$ \% in the 9 years following
the all-time high of 31 Dec. 1989. In figure \ref{decnikkei}, we see (the
logarithm of)  the Nikkei from 31 Dec. 1989
until 31 Dec. 1998. The 3 fits are equations (\ref{1feq}), (\ref{2feq}) and
(\ref{3feq}) respectively \cite{antibulle}\,:
\bea
\log\lp p\lp t\rp \rp &\approx& A_1 + B_1 \tau ^{\beta}
+C_1 \tau^\beta \cos\left[ \omega_1 \log \lp \tau \rp +\phi_1 \right]
\label{1feq} \\
\log\lp p(t) \rp &\approx& A_2 + \frac{\tau^\beta}{\sqrt{1+\left(\frac{\tau}
{T_1}\right)^{2\beta}}} \left\{B_2+ C_2\cos\left[\omega_1\log \tau +
\frac{\omega_2}{2\beta}\log \left(1+\left(\frac{\tau}
{T_1}\right)^{2\beta}\right)+\phi_2\right]\right\}, \label{2feq}
\eea
$$
\log\lp p(t) \rp \approx A_3 + \frac{\tau^\beta}{\sqrt{1+\left(\frac{\tau}
{T_1}\right)^{2\beta} + \left(\frac{\tau}{T_2}\right)^{4\beta}}}
$$
\be \label{3feq}
\left\{B_3+ C_3\cos\left[\omega_1\log \tau +
\frac{\omega_2}{2\beta}\log\left(1+\left(\frac{\tau}
{T_1}\right)^{2\beta}\right)+
\frac{\omega_3}{4\beta}\log\left(1+\left(\frac{\tau}
{T_2}\right)^{4\beta}\right) + \phi_3\right]\right\}~,
\ee
where $\tau = t - t_c $.
Equation (\ref{3feq}) predicts the transition from the log-frequency 
$\omega_1$ close to $t_c$ to $\omega_1 + \omega_2$ for $T_1 < \tau < T_2$ and 
to the log-frequency $\omega_1 + \omega_2 + \omega_3$ for $T_2
< \tau$. The equations (\ref{2feq}) (resp.
(\ref{3feq})) extend the renormalisation group approach
to the second (resp. third) order of perturbation \cite{antibulle,SJ97}.

The interval used for equation (\ref{1feq}) is until
mid-1992 and for equation (\ref{2feq}) until mid-1995. The
parameter values
produced by the different fits with equations (\ref{1feq}) and (\ref{2feq})
agree remarkably well and the values for the exponent $\beta$ also
agree well as shown in table \ref{table2}.  The
fit with (\ref{3feq}) was, due to the large number of free variables,
performed differently and the parameter values for $t_c$, $\beta$, and
$\omega_1$ determined from the fit with (\ref{2feq}) was kept fixed and only
$T_1$, $T_2$, $\omega_2$, $\omega_3$ and $\phi_3$
where allowed to adjust freely. What lends credibility to the fit with
equation (\ref{3feq}) is that
despite it complex form, we get values for the two cross-over time scales
$T_1$, $T_2$ which correspond very nicely to what is expected from
the theory: $T_1$ has moved  down to $\approx 4.4$ years in agreement
with the time interval used for equation (\ref{2feq}) and $T_2$ is
$\approx 7.8$ years, which is compatible with the 9 year interval of the
fit. This does not mean that fitting is not very degenerate, it is, but the
ranking of $T_1$ and $T_2$ is always the same and the values
given do not deviate much from the ones in the caption of figure
\ref{decnikkei}, i.e., by $\pm$ one year.

The value $\omega_1 \approx 4.9$ correspond to $\lambda \approx 3.6$, which
is not in the range found for the bubbles, see table \ref{table1}. An
additional difference between the anti-bubble in the Nikkei and these other
cases is the strength of the oscillations compared to the leading behaviour.
For the Nikkei, it is $\approx 20$\%, {\it i.e.}, $2-4$ times as large as the
amplitude obtained for the stock market and Forex bubbles.

\subsection{Log-periodic analysis of the Russian stock market} \label{rus}

Intrigued by the claims of K. Ilinski \cite{Ilinski} of a genuine stock
market bubble, followed by a crash, {\it without} log-periodic signatures
on the Russian stock market, we have analysed 4 Russian stock market
indices from their start up till present. The 4 indices was The Russian
Trading System Interfax Index (IRTS), The Agence Skate Press Moscow
Times Index (ASPMT), The Agence Skate Press General Index (ASPGEN) and
The Credit Suisse First Boston Russia Index (ROSI). The ROSI is generally
considered the best of the 4 and we will put emphasis on the results
obtained with that index. As we shall return to later, the Russian stock
market is highly volatile, which means that great care must be taken in
maintaining a representative stock market index. This is the primary
reason for using 4 indices in the analysis.

\begin{figure}
\begin{center}
\epsfig{file=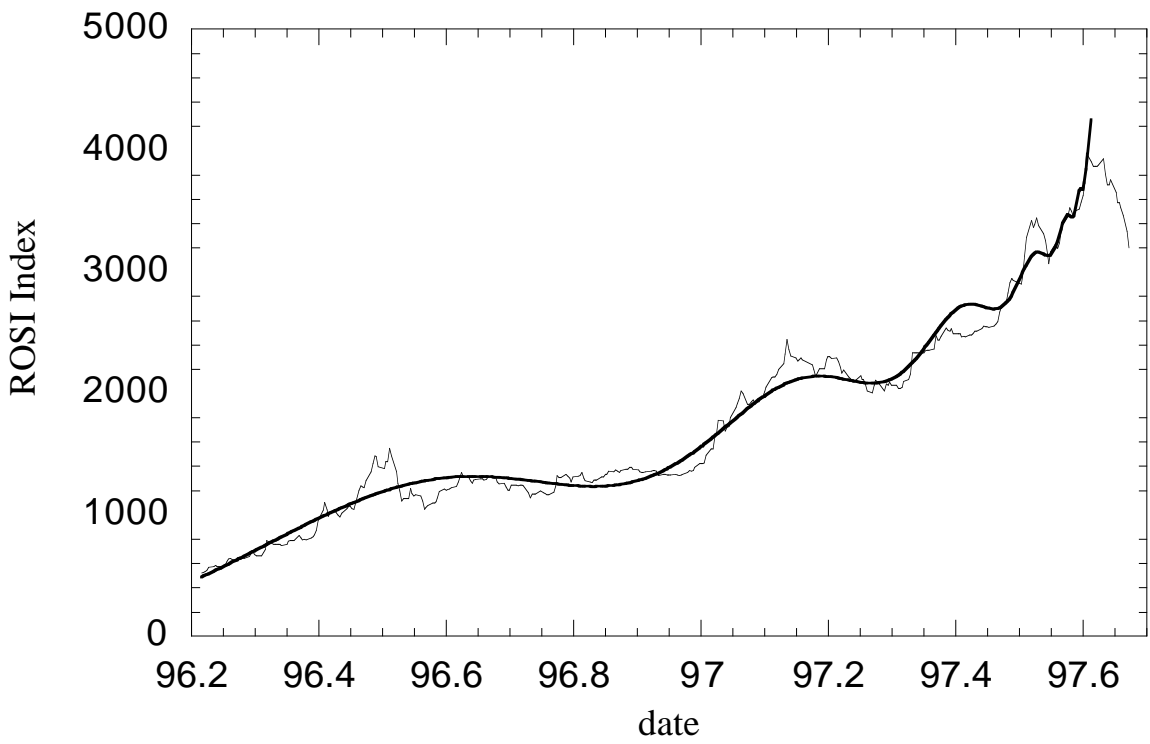}
\caption{\protect\label{acc} The ROSI Index fitted with equation
\protect\ref{lppow} over the interval shown. The parameter values of the fit
with equation (\protect\ref{lppow})
are $A_1\approx 4254 , B_1\approx -3166 , C_1\approx  246 , \beta \approx
0.40 , t_c \approx  97.61 , \phi_1 \approx  0.44 , \omega_1 \approx  7.7$.  }
\end{center}
\end{figure}

In figure \ref{acc}, we see the ROSI fitted with equation (\ref{lppow})
in the interval $\left[96.21:97.61\right]$. The interval is chosen by
identifying the start of the bubble and the end represented by the date
of the highest value of the index before the crash similarly to the $7$
major market crashes discussed previously. For all 4 indices, the
same start- and end-day could be identified $\pm 1$ day.

\begin{figure}
\begin{center}
\epsfig{file=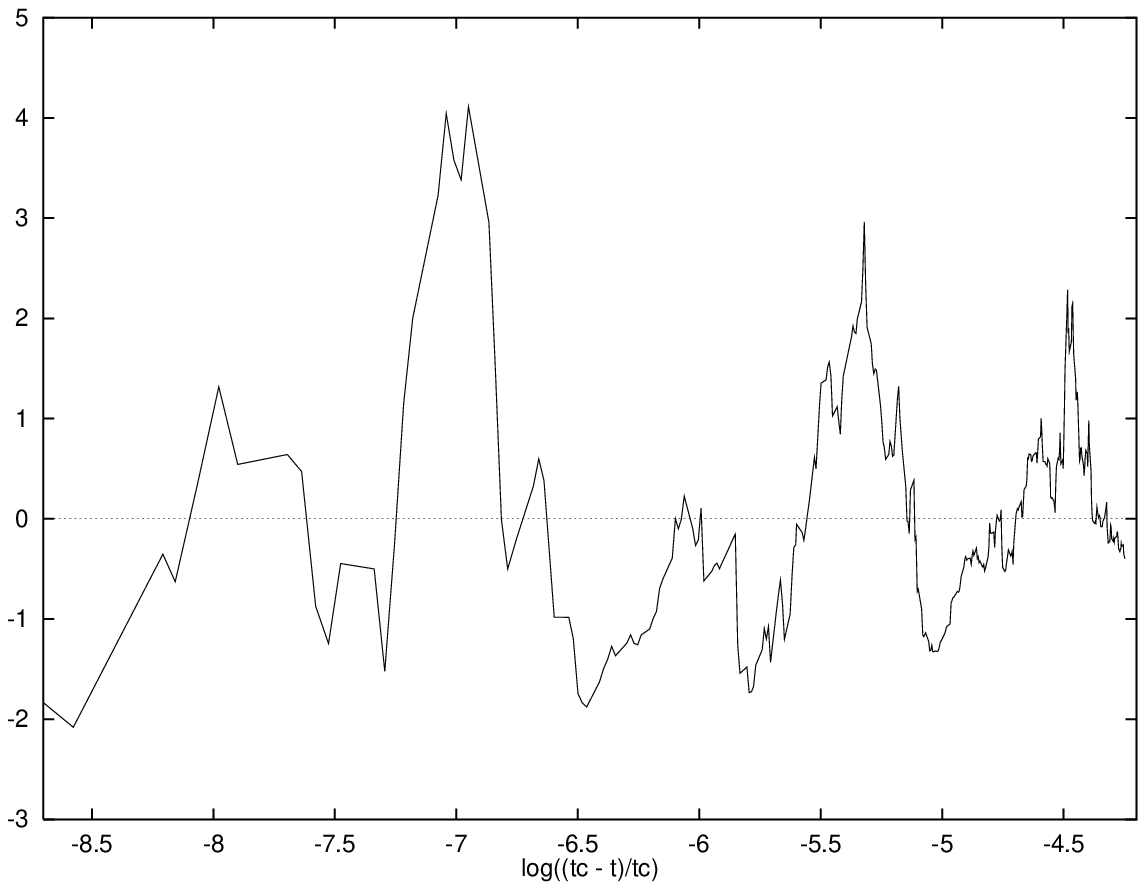,  width=10cm}
\caption{\protect\label{resirosiacc} The residual as defined by the
transformation (\protect\ref{residue}) as a function of
$\log\lp \frac{t_c - t}{t_c}\rp$ for the Russian data shown in figure
\protect\ref{acc}.}
\end{center}
\end{figure}

\begin{figure}
\begin{center}
\epsfig{file=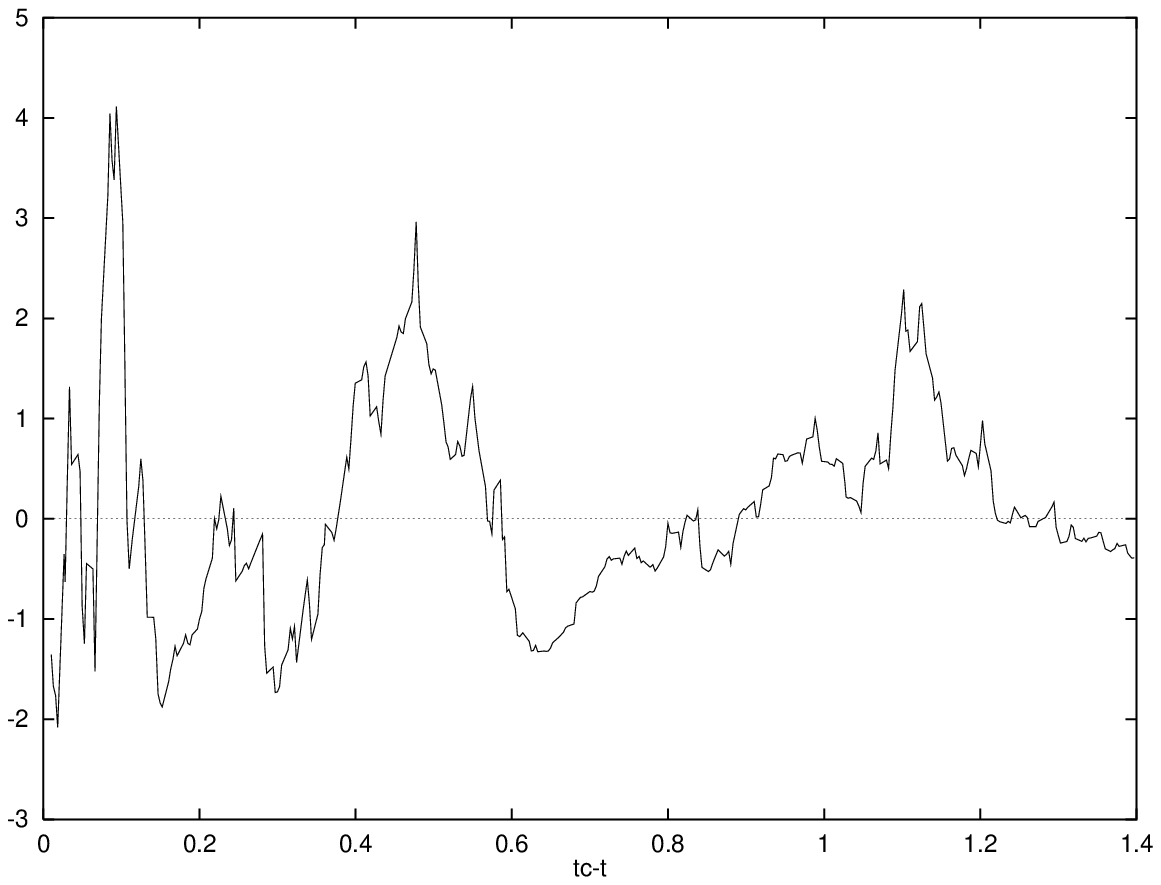,  width=10cm}
\caption{\protect\label{resirosiaccdd} The residual as defined by the
transformation (\protect\ref{residue}) as a function of $t_c -t$ for the
Russian data shown in figure \protect\ref{acc}.}
\end{center}
\end{figure}

In figures \ref{resirosiacc} and \ref{resirosiaccdd}, we see
the detrended data using the transformation (\ref{residue}). The
conclusions made in relation to the corresponding figures for the 1987
Wall street crash (figures \ref{residue87} and \ref{linresidue87})
are again valid supporting both an over-all power law rise
as well as log-periodic signatures for the Russian crash of 1997.

As can be seen from table \ref{table2}, the non-dimensional parameters $\beta$,
$\omega_1$ and $\lambda$ as well as the predicted time of the crash $t_c$ for
the fit to the different indices agree very well except for the exponent
$\beta$ obtained from the ASPGEN Index. In fact, the value obtained for the
prefered scaling ratio $\lambda$ is fluctuating by no more than $5$\% for
the 4 fits showing
numerical stability.

\begin{table}[h]
\begin{center}
\begin{tabular}{|c|c|c|c|c|c|c|c|c|c|c|c|} \hline
Bubble&$t_c$& $t_{max}$ & $t_{min}$ & drop & $\beta$& $\omega_1$ &
$\lambda$ & $A$ & $B$ & $C$ & $Var$ \\ \hline
ASPMT& $97.61$& $97.61$ & $97.67$ & $17\%$ & $0.37$   & $7.5$    & $2.3$
& $1280$ & $-1025$ & $59.5$ & $907$ \\ \hline
IRTS  & $97.61$& $97.61$ & $97.67$ & $17\%$ & $0.39$   & $7.6$    &
$2.3$ & $633$ & $-483$ & $38.8$ & $310$ \\ \hline
ROSI  & $97.61$ &  $97.61$ & $97.67$ & $20\%$ & $0.40$   & $7.7$    &
$2.3$ & $4254$ & $-3166$ & $246$ & $12437$ \\ \hline
ASPGEN & $97.62$ & $97.60$ & $97.67$ & $8.9\%$ & $0.25$   & $8.0$    &
$2.2$ & $2715$ & $-2321$ & $72.1$ & $1940$ \\ \hline
Anti-bubble & $t_c$  &  $t_{max}$ & $t_{min}$ & drop &$\beta$ &
$\omega_1$ & $\lambda$  & $A$ & $B$ & $C$ & $Var$\\ \hline
ROSI & $97.72$ & $97.77$ & $98.52$ & $74\%$ & $0.32$   & $7.9$    &
$2.2$ & $4922$ & $-3449$ & $472$ & $59891$ \\ \hline
Nikkei eq.(\protect\ref{1feq}) & $89.99$ & $90.00$ & $92.63$ & $63\%$ &
$0.47$ & $4.9$ & $3.6$ & $10.7$ & $-0.54$ & $-0.11$ & $0.0029$\\ \hline
Nikkei eq.(\protect\ref{2feq})& $89.97$ & $90.00$ & $95.51$ & $63\%$ &
$0.41$ & $4.8$ & $3.7$ & $10.8$ & $-0.70$ & $-0.11$ & $0.0600$  \\ \hline
\end{tabular}
\end{center}
\caption{\label{table2} $t_c$ is the predicted time of the crash from
the fit of the market index to the equation (\protect\ref{lppow}). The
other parameters of the fits to the preceding bubble are also given.
The error $Var$ is the variance between the data and the fit, equation 
(\protect\ref{var}), and has units $price^2$ except for the Nikkei, where
the units are $[\log(price)]^2$. The fit to the bubble is performed up to the
time at which the market index achieved its highest maximum before the crash.
The parameters $t_c$, $\beta $, $\omega_1 $ and $\lambda $ correspond to
the fit with equation (\protect\ref{lppow}), where $t_c$ and $t$ has been
interchanged. Here $t_{max}$ and $t_{min}$ represent the endpoints of
the interval fitted.}

\end{table}

The origin of this bubble is well-known. In 1996 large international
investors (read US, German and Japanese) began to invest heavily in
the Russian markets believing that the financial situation of Russia
had finally stabilised. Nothing was further from the truth
\cite{Intriligator,Malki} but the belief and hope in a new investment haven
with large returns led to herding and bubble development.
This means that the same herding, which created the log-periodic bubbles on
Wall Street (1929, 1987, 1998), Hong Kong (1997) and the Forex (1985, 1998),
entered an emerging market and brought along the same log-periodic trends
characterising the global markets. The fact that the consistent values of
$\lambda$ obtained for the 4 indices of the Russian market are comparable to
that of the Wall Street, Hong Kong and Forex crashes supports this
interpretation.
Furthermore, it supports the idea of the stock market as a self-organising
complex system of surprising robustness.

Inspired by this clear evidence of log-periodic oscillations decorating the
power law acceleration signaling a bubble in the Russian stock market, we
extended our analysis and search for possible log-periodic signatures in the
``anti-bubble" that followed the log-periodic bubble described above.

\begin{figure}
\begin{center}
\epsfig{file=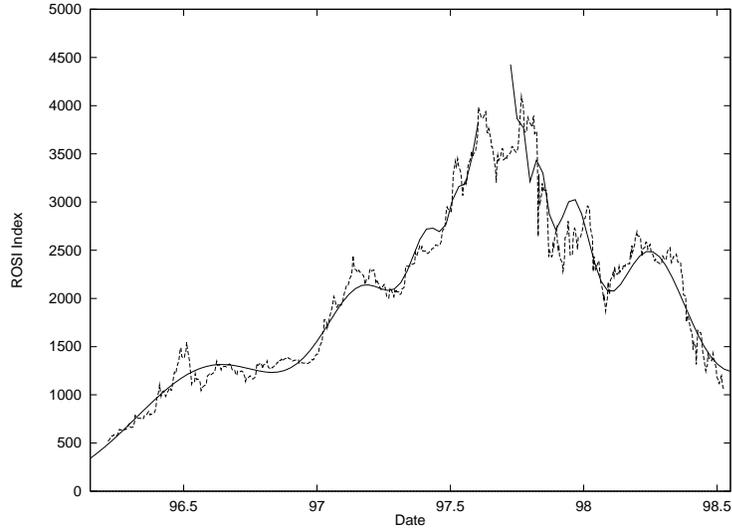,  width=10cm}
\caption{\protect\label{accdec} Symmetric ``bubble'' and ``anti-bubble''\,:
in addition to the ascending part of the ROSI Index which is reproduced
from figure \ref{acc} with the same fit, we show the deflating part
fitted with equation \protect\ref{1feq}.
The parameter values of the fit
are $A_1\approx 4922 , B_1\approx -3449 , C_1\approx  472 , \beta \approx
0.32 , t_c \approx  97.72 , \phi_1 \approx  1.4 , \omega_1 \approx  7.9$.  }
\end{center}
\end{figure}

As described in section \ref{decay},
the recent decay of the Japanese Nikkei index starting 1. Jan 1990
until present can be excellently
model by a log-periodically decorated power law decay. In figure \ref{accdec},
we show the ROSI index for the  ``anti-bubble'', signaling the collapse of
the Russian stock market, fitted with equation (\ref{lppow}) where
$t_c$ and $t$ has been interchanged. We use the price and not the logarithm of
the price to keep the symmetry with the description of the bubble and from
the consideration of the relatively short time scales involved.

In table \ref{table2}, the corresponding values for the fit of the anti-bubble
with equation (\ref{1feq}) listed. The agreement between the bubble and the
anti-bubble with respect to the values of the non-dimensional variables
$\beta$, $\omega_1$ and $\lambda$ as well as
$t_c$ is surprisingly good. However, the numerical stability of the result
for the anti-bubble cannot be compared with that of the preceding bubble and
depends on the index as well as the endpoint of the interval used in the
fitting. However, the ``symmetry'' around $t_c$ is rather striking
considering the nature of the data and quite visible to the naked eye.

There are (at least) two reasons for the numerical instability of the fit
to the anti-bubble. First, as an emerging market in decay the Russian stock
market is too volatile for a smooth function as a cosine and we
are currently investigating alternative parametrisations \cite{inpreparation}.
Second, the ``noise'' in the Russian indices were very large in that period,
in particular due
to the ``depletion'' of stocks. This is clearly illustrated by the heavy
rearrangement that the ASPGEN and related indices went through following
Aug.~17, 1998.

\begin{figure}
\begin{center}
\epsfig{file=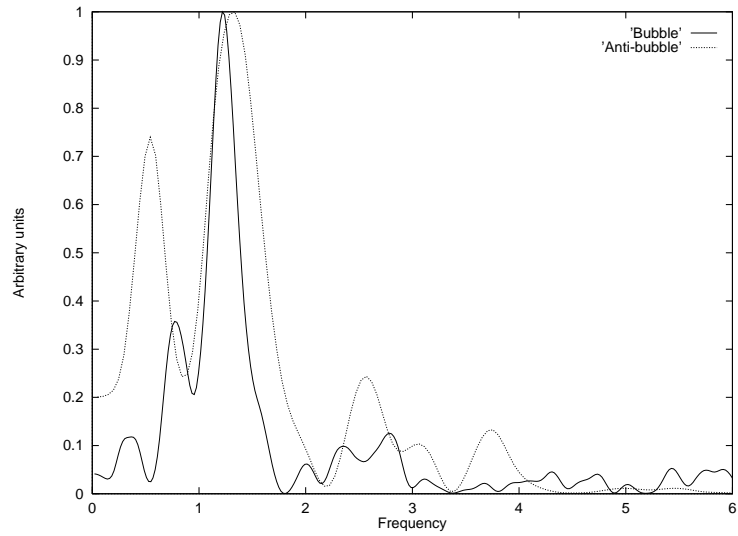,width=10cm}
\caption{\protect\label{rosiaccdecfp}The Lomb periodogram for the bubble and
anti-bubble shown in figure \protect\ref{accdec}.
The significance of the peak should be estimated against the noise level.}
\end{center}
\end{figure}

In figure  \ref{rosiaccdecfp}, we see the spectral analysis for the bubble
and anti-bubble. Whereas the agreement between the periodogram and the fit
with equation (\ref{lppow}) is excellent (within $1\%$), the deviation for
the anti-bubble is $\approx 5\%$. Furthermore, the residual for the anti-bubble
used in the frequency analysis shown in figure \ref{rosiaccdecfp} was truncated
by a few weeks due to an increasing deviation between data and fit.

It may seem odd to argue for the log-periodic oscillations while one can
forcefully argue that the market is largely reflecting the vagaries of the
Russian political institutions. For instance, in the anti-bubble case,
Feb--April 1998 was a revival period for the market characterized by the
returning of western investors after the post-crash calm-down. This can be
followed by studying the dynamics of the Russian external reserves. The timing
of the return can be argued to be dictated by the risk policies of larger
investors more than  anything else. The next large drop of the Russian index
in April 1998 originated by the decision of Mr. Yeltsin to sack Mr.
Chernomyrdin's government, which destabilized the political situation
and created uncertainty. Further political disturbance was introduced twice
by the Duma when it rejected  Mr. Yeltsin's candidates for the prime minister
office and put itself on the brink of dissolution. Opposed to this, we argue
that one must not mistake a global unstable situation for the specific
historical action that triggered the instability. Consider a ruler put
vertically on a table. Being in an unstable position, the stick will fall in
some direction and the specific air current or slight initial imperfection in
the initial condition are of no real importance. What {\it is} important is
the intrinsically unstable initial state of the
stick. We argue that a similar situation applies for crashes. They occur
because the market has reached a state of global instability. Of course, there
will always be specific events which may be identified as triggers of market
motions but they will be the revelators rather than the deep sources of the
instability. Furthermore, the political events must also be considered as
revelators of the state of the dynamical system which includes the market.
There is, in principle, no decoupling between the different events.
Specifically, the Russian crash may have been triggered by the Asian crises,
but it was to a large extent fueled by the collapse of a banking system, which
in the course of the bubble had created an outstanding debt of $\$ 19.2$ 
billion \cite{Malki}.

\section{Conclusion}

We have presented a synthesis of the available empirical evidence in the
light of recent theoretical developments for the existence of characteristic
log-periodic signatures of growing bubbles in a variety of stock markets as
well as currencies. We have here documented 8 unrelated crashes from 1929 to
1998, on stock markets as diverse as the US, Hong-Kong or the Russian market
and on currencies. In addition, we have discovered a significant bubble on Wall
Street ending in 1962 \cite{JLS} as well as ``anti-bubbles''on the Nikkei since
1990 and the Gold (after the 1980 bubble maximum) \cite{antibulle}. Quite
unexpectedly, we have shown that the Russian bubble crashing in Aug. 1997
had close to identical power law and log-periodic behaviour to the bubbles
observed on Wall Street, the Hong-Kong stock market and on currencies. To our
knowledge, no major financial crash preceded by an extended bubble
has occurred in the past 2 decades without exhibit a log-periodic
signature. In this context, note that the novel analysis of the
Russian index presented here was motivated by Ilinski's claim of a crash
without log-periodic signature, which we have shown to be incorrect.

All these results, taken together with the remarkable robustness and 
consistency of the estimation of the exponent $\beta$ as well as the 
more important statistics the scaling ratio $\lambda$, make the case for 
power law acceleration and log-periodicity very strong. In our opinion, one 
can no more ignore theese very specific and strong signatures which is 
characteristic of developing bubbles and this calls for further investigations 
to unravel in more depths the underlying economical, financial and behavioural 
mechanisms.

These different cases, together with the 1962 ``slow event'' as well as the
``anti-bubbles'',
show that the log-periodic critical theory applies both to bubbles ending
in a sudden
crash as well as to bubbles landing smoothly. This is in fact a strong
prediction of our rational model of imitative behaviour.

What we have attempted here is not to explain why crashes happens or bubbles
exists, but to quantify the process taking place during extended bubbles that
very often lead to the rapid regime-switching a crash represent.
We have offered evidence that bubbles and anti-bubbles
have log-periodic and power law characteristics and hence provides for
a quantification of extended ``moods'' on the markets.
We have furthermore,
provided a model which contains the observed signatures. Whether, as proposed,
discrete scale invariance or some other mechanisms \cite{Revue} are
responsible for the prominent log-periodic signatures observed, only a more
detailed microscopic model can answer.

\vskip 0.5cm
\noindent {\bf Acknowledgement}:
The authors are grateful to K. Ilinski for bringing their attention
to the Russian stock market and E. de Malherbes for providing the Russian data.

\newpage

\end{document}